\documentstyle[epsf,12pt,prd,aps]{revtex}

\begin{document}

\def\R{{\bf R}}
\def\Z{{\bf Z}}
\def\x{{\bf x}}
\def\del{\partial}
\def\Lap{\bigtriangleup}
\def\Div{{\rm div}\ }
\def\rot{{\rm rot}\ }
\def\curl{{\rm curl}\ }
\def\grad{{\rm grad}\ }
\def\Tr{{\rm Tr}}
\def\^{\wedge}
\def\goinf{\rightarrow\infty}
\def\goes{\rightarrow}
\def\bm{\boldmath}
\def\-{{-1}}
\def\inv{^{-1}}
\def\sqr{^{1/2}}
\def\isqr{^{-1/2}}

\def\reff#1{(\ref{#1})}
\def\vb#1{{\partial \over \partial #1}} 
\def\Del#1#2{{\partial #1 \over \partial #2}}
\def\Dell#1#2{{\partial^2 #1 \over \partial {#2}^2}}
\def\Dif#1#2{{d #1 \over d #2}}
\def\Lie#1{ {\cal L}_{#1} }
\def\diag#1{{\rm diag}(#1)}
\def\abs#1{\left | #1 \right |}
\def\rcp#1{{1\over #1}}
\def\paren#1{\left( #1 \right)}
\def\brace#1{\left\{ #1 \right\}}
\def\bra#1{\left[ #1 \right]}
\def\angl#1{\left\langle #1 \right\rangle}
\def\lvector#1#2#3#4{\paren{\begin{array}{c} #1 \\ #2 \\ #3 \\ #4 \end{array}}}
\def\vector#1#2#3{\paren{\begin{array}{c} #1 \\ #2 \\ #3 \end{array}}}
\def\svector#1#2{\paren{\begin{array}{c} #1 \\ #2 \end{array}}}
\def\matrix#1#2#3#4#5#6#7#8#9{
        \left( \begin{array}{ccc}
                        #1 & #2 & #3 \\ #4 & #5 & #6 \\ #7 & #8 & #9
        \end{array}     \right) }
\def\smatrix#1#2#3#4{
        \left( \begin{array}{cc} #1 & #2 \\     #3 & #4 \end{array}     \right) }

\def\GL#1{{\rm GL}(#1)}
\def\SL#1{{\rm SL}(#1)}
\def\PSL#1{{\rm PSL}(#1)}
\def\O#1{{\rm O}(#1)}
\def\SO#1{{\rm SO}(#1)}
\def\IO#1{{\rm IO}(#1)}
\def\ISO#1{{\rm ISO}(#1)}
\def\U#1{{\rm U}(#1)}
\def\SU#1{{\rm SU}(#1)}

\def\diffeos{diffeomorphisms}
\def\diffeo{diffeomorphism}
\def\Teich{{Teichm\"{u}ller }}
\def\Poin{{Poincar\'{e} }}

\def\Gam{\mbox{$\Gamma$}}
\def\d{{\rm d}}
\def\n{\mbox{$n$}}
\def\sz{\mbox{VI${}_0$}}
\def\svz{\mbox{VII${}_0$}}
\def\tR{{}^{(3)}\! R}
\def\fR{{}^{(4)}\! R}
\def\Isom{{\rm Isom}}
\def\Esom{{\rm Esom}}
\def\SL{\widetilde{{\rm SL}}(2,\R)}
\def\SLx{{\rm SL}(2,\R)}
\def\Nil{{\rm Nil}}
\def\Sol{{\rm Sol}}
\def\Lapp{\bar g}

\def\hh{{h}}
\def\gg{{\rm g}}
\def\uh#1#2{\hh^{#1#2}}
\def\dh#1#2{\hh_{#1#2}}
\def\ug#1#2{\gg^{#1#2}}
\def\dg#1#2{\gg_{#1#2}}
\def\uug#1#2{\tilde{\gg}^{#1#2}}
\def\udg#1#2{\tilde{\gg}_{#1#2}}
\def\udh#1#2{\tilde{\hh}_{#1#2}}
\def\ustd#1#2{\tilde{h}_{#1#2}^{\rm std}}

\def\BVII#1{\mbox{Bianchi VII${}_#1$} }
\def\BVI#1{\mbox{Bianchi VI${}_#1$} }
\def\four#1{{}^{(4)}\! #1}

\def\MM{\four{M}}
\def\uMM{\four{\tilde{M}}}
\def\Mtil{\tilde M}

\def\Mt{\mbox{$M_t$}}
\def\uMt{\mbox{$\tilde{M}_t$}}
\def\uMi{\mbox{$\tilde{M}_{t_0}$}}
\def\UC{\mbox{$(\uMM,\udg ab)$}}
\def\CU{\mbox{$(\MM,\dg ab)$}}
\def\UCh{\mbox{$(\uMt,\udh ab)$}}
\def\UCs{\mbox{$(\Mtil,\udh ab)$}}
\def\UChi{\mbox{$(\uMti,\udh ab)$}}
\def\CUh{\mbox{$(\Mt,\dh ab)$}}
\def\Isomf{\mbox{$\Esom\uMt$}}
\def\Isomff{\mbox{$\Esom(\uMt,\uMM)$}}
\def\dataset{$(\dh ab, K_{ab})$}
\def\onedata{$(\dh \mu\nu, K_{\mu\nu})$}

\def\vars{F}
\def\pps{P}
\def\tei{\mbox{\bm $\tau$}}
\def\cur{\mbox{\bm $r$}}
\def\tcv{\mbox{$(\cur, \tei, v)$}}
\def\u{{\bf u}}
\def\g{{\bf g}}
\def\h{{\bf h}}
\def\std{{\rm std}}
\def\dyn{{\rm dyn}}
\def\UCx#1{(\uMM,\udg ab{#1})}
\def\CUx#1{(\MM,\dg ab{#1})}
\def\gdy#1{\udg ab^\dyn{#1}}
\def\UCgdyx#1{(\uMM,\gdy{#1})}
\def\UCu{\mbox{$\UCx{[\u]}$}}
\def\CUug{\mbox{$\CUx{[\u,\g]}$}}
\def\UCgdycurtei{\mbox{$\UCgdyx{(\cur,\tei)}$}}
\def\hst#1{\udh ab^\std{#1}}
\def\hdy#1{\udh ab^\dyn{#1}}
\def\UChx#1{(\uMt,\udh ab{#1})}
\def\UCstx#1{(\Mtil,\hst{#1})}
\def\UCdyx#1{(\Mtil,\hdy{#1})}
\def\UChdyx#1{(\uMt,\hdy{#1})}
\def\UChu{\mbox{$\UChx{[\u]}$}}
\def\UCst{\mbox{$\UCstx{}$}}
\def\UCstcur{\mbox{$\UCstx{[\cur]}$}}
\def\UCdycurtei{\mbox{$\UCdyx{[\cur,\tei]}$}}
\def\UChdycurtei{\mbox{$\UChdyx{[\cur,\tei]}$}}

\def\s#1{\sigma^{#1}}
\def\a#1#2{a_{#1}{}^{#2}}
\def\p#1#2{p^{#1}{}_{#2}}
\def\dug#1#2{g_{#1}{}^{#2}}
\def\H#1#2{H_{#1#2}}
\def\f#1#2{f^{#1}{}_{#2}}

\def\wa{&=&}
\def\wb{&\equiv&}

\def\eqnI#1{(#1)${}_{\rm I}$}
\def\eqnII#1{(#1)${}_{\rm II}$}

\def\ack{\section*{Acknowledgments}
M.T. acknowledges financial support from the Japan Society for the Promotion
of Science and the Ministry of Education, Science and Culture.}

\def\abstract#1{\begin{center}{\bf ABSTRACT}\end{center}
\par #1}
\def\title#1{\begin{center}{\large {#1}}\end{center}}
\def\author#1{\begin{center}{\sc #1}\end{center}}
\def\address#1{\begin{center}{\it #1}\end{center}}

\def\pubnum{97--24}

\begin{titlepage}
\hfill
\parbox{6cm}{{YITP--\pubnum} \par May 1997}
\par
\vspace{1.5cm}
\begin{center}
\Large
Hamiltonian structures for compact homogeneous universes
\end{center}
\vskip 2cm
\author{Masayuki TANIMOTO${}^\dagger$\footnote{JSPS Research Fellow},
Tatsuhiko KOIKE${}^\flat$ and Akio HOSOYA${}^\natural$
}
\address{$\dagger$ Yukawa Institute for Theoretical Physics,
        Kyoto University, Kyoto 606-01, Japan. \\
Electronic mail: tanimoto@yukawa.kyoto-u.ac.jp \\
$\flat$ Department of Physics, Keio University, Kanagawa 230, Japan. \\
Electronic mail: koike@rk.phys.keio.ac.jp \\
$\natural$ Department of Physics, Tokyo Institute of Technology,
Tokyo 152, Japan. \\
Electronic mail: ahosoya@th.phys.titech.ac.jp
}
\vskip 1 cm

\abstract{ Hamiltonian structures for spatially compact locally
  homogeneous vacuum universes are investigated, provided that the set
  of dynamical variables contains the \Teich parameters, parameterizing
  the purely global geometry. One of the key ingredients of our
  arguments is a suitable mathematical expression for quotient
  manifolds, where the universal cover metric carries all the degrees of
  freedom of geometrical variations, i.e., the covering group is fixed.
  We discuss general problems concerned with the use of this expression
  in the context of general relativity, and demonstrate the reduction of
  the Hamiltonians for some examples. For our models, all the dynamical
  degrees of freedom in Hamiltonian view are unambiguously interpretable
  as geometrical deformations, in contrast to the conventional open
  models.}
\end{titlepage}

\addtocounter{page}{1}

\section{Introduction}

Homogeneous cosmological models \cite{LL,RS,Wa} have been served as a
good arena for, e.g., quantum cosmology (e.g., \cite{Hal,D'Ea}), as well
as observational cosmology (e.g. \cite{Wein}), after first investigated
in connection with the singularity problem \cite{BKL}. As for
Hamiltonian structures of them, however, there are controversies. For
example, it is well known that the models known as Bianchi class B do
not possess a natural Hamiltonian reduced from the full Hamiltonian
(See, e.g., Ref.\cite{RS}, p.193).  Even for the class A models, a sort
of discrepancies of dynamical degrees of freedom is pointed out by
Ashtekar and Samuel\cite{AS}. For example, the Kasner solution, which is
the vacuum solution of Bianchi I, gives one for the number of dynamical
degrees of freedom, i.e., the number of free parameters which can be
specified freely at an initial Cauchy surface.  Odd number of dynamical
degrees of freedom, however, cannot come out from a Hamiltonian system.
So, when a Hamiltonian is needful, people usually work with the diagonal
model, which has three dynamical variables and gives four dynamical
degrees of freedom in the Hamiltonian view.  Moreover, if we work with
the full, not diagonal model, which may be the most natural in the
Hamiltonian view, we have ten dynamical degrees of freedom with six
dynamical variables. Thus, three numbers (i.e., 1, 4 and 10) of
dynamical degrees of freedom are possible for Bianchi I!

However, this discrepancy is for the conventional {\it open} model.  It
is a recent progress \cite{KTH,TKH} that a satisfactory framework has
been established for the construction of spatially {\it compact} locally
homogeneous spacetimes. (We shall hereafter refer to Refs.\cite{KTH} and
\cite{TKH} as I and II, respectively.) We can now also unambiguously
count the dynamical degrees of freedom in the above sense.  In this
framework, we ``compactify'' a usual spatially open homogeneous model by
identifying spatial points appropriately. So, the new parameters come
into the model which parameterize the identifications. An important
point in dynamical view is that this compactification gives rise to new
degrees of freedom of spatial deformations, known in mathematics as
\Teich deformations \cite{Oh,KLR}.  \Teich deformations obviously do not
vary any local geometries.  We call the parameters which parameterize
the \Teich deformations {\it \Teich parameters}, and denote them
collectively as \tei.  The natural dynamical variables of a spatially
compact homogeneous (SCH) spacetime therefore consist of \tcv, where
\cur\ are the (local) curvature parameters, and $v$ is the total spatial
volume.  (For definiteness, we consider a vacuum case.)  We denote the
space spanned by \tcv\ and the cotangent bundle of it as $\vars$ and
$\pps=T_*\vars$, respectively, and call $\pps$ the pseudo-phase space.
What we shall investigate in this paper is whether $\pps$ admits a
Hamiltonian structure, so in other words, whether it is good enough to
call it a phase space.

We shall concentrate on the natural Hamiltonian structures, i.e., we
take interests only in the phase spaces reduced from the full phase
space which is made from the Einstein-Hilbert action.  Thus, all the
dynamical parameters \tcv\ should appear as the metric components.  Note
that in our approach in II, parameters \tcv\ appeared both in the
universal cover metric and the covering map --- that is, the dynamical
degrees of freedom were distributed in the two parts.  To put them
together into the universal cover metric, we need to be able to fix the
covering map in an appropriate sense.  In other words, all the dynamical
degrees of freedom of the compact model need to be able to be expressed
in the universal cover metric with a fixed covering map.  Though at
first sight this seems always possible, we should note that the metric
obtained in this way, which has finite degrees of freedom of variation,
does not always define a consistent cut in the full phase space.  (This
is the same situation as that for a partial differential equation, an
arbitrary ansatz of solution does not lead to a true solution.)  We will
show for which models we can do this, and show why we cannot for the
models we cannot.

The organization of this paper is as follows.  In section \ref{sec2}, we
review the previous work briefly, which gives a basis of our argument in
the subsequent sections.  In section \ref{sec3}, we show how we can fix
the covering map and thereby the universal cover metric can carry all
the dynamical degrees of freedom.  A class of \diffeos\ (TDs) and a
subclass of it (HPTDs) are introduced.  Section \ref{sec4} is devoted to
explicit examples.  Section \ref{lastsec} is for conclusions and
remarks.

As we have already done, we abbreviate the words {\it spatially compact
  homogeneous} (or more precisely, {\it spatially compact and spatially
  locally homogeneous}) {\it spacetimes} as {\it SCH spacetimes}.  Words
in the title {\it compact homogeneous universes} are equivalent to the
words {\it SCH spacetimes}.

\section{Construction and dynamical aspects of compact homogeneous universes}
\label{sec2}

In this section, we briefly review the previous work, which concerns the
construction of SCH spacetimes, and the dynamical aspects of them,
namely the time-developments of the dynamical variables and the
dynamical degrees of freedom.  We refer the reader to II for explicit
examples.

First, let us summarize the construction of the smooth set of compact
homogeneous universes as a natural dynamical system.  We assume that the
topology of the compact homogeneous universes we consider is
$M\times\R$, where $M$ is a spatial manifold and $\R$ corresponds to
time.  The fundamental group $\pi_1(M)$ of the space is supposed to be
fixed in what follows.

Consider a spatially homogeneous spacetime \UC\ and a homogeneous
spatial section \UCh\ of \UC. We denote the isometry groups of them as
$\Isom\uMM$ and $\Isom\uMt$, respectively.  [Here, the subscript $t$
parameterizes the spatial sections of \UC, but we only consider a fixed
value of $t$ for a while, so that $t$ is just a reminder for \uMt\ being
a section of $\uMM$.]  Note that the subgroup of $\Isom\uMM$ which
preserves \uMt\ can be identified with a subgroup of $\Isom\uMt$. We
call this subgroup the {\it extendible isometry group} of \UCh, and
denote it as \Isomf.  To compactify \UCh, we embed $\pi_1(M)$ {\it not}
into $\Isom\uMt$ but into \Isomf, since by doing so, we can obtain a
total smooth SCH spacetime \CU.  This is simply because \Isomf\ and
therefore the embedding $\Gam\subset\Isomf$ are also subgroups of
$\Isom\uMM$, so that the quotient $\UC/\Gam=\CU$ is guaranteed to be
smooth.

This construction of a compact homogeneous universe tells us what {\it
  define} the natural smooth set of compact homogeneous universes as our
dynamical system.  We think they are the fundamental group of the
spatial section $\pi_1(M)$ {\it and} the group structure of \Isomf.  We
therefore fix the two groups, as well as an Einstein's equation, then
think of the set $C$ of all possible compact homogeneous universes
prescribed by them as a candidate of the natural smooth set.  The set
$C$ is, more explicitly, defined as the smooth set of all possible pairs
$(u,\Gam)$ of a universal covering spacetime $u$ and a covering group
$\Gam$ such that --- (1) Each $u$ has a spatial section admitting the
same extendible isometry group \Isomf. (2) Each $u$ satisfies the same
Einstein's equation. (3) Each $\Gam$ acting on $u$ is a possible
embedding of $\pi_1(M)$ into \Isomf.  We also define $U$ as the smooth
set of universal covers which satisfy the conditions (1) and (2).

Each element $(u,\Gam)$ in $C$ is naturally identified with the compact
homogeneous universe $u/\Gam$.  Our natural smooth set $\bar C$ of
compact homogeneous universes shall be obtained by identifying the
isometric elements in $C$.  To do this, we summarize key facts on
\diffeos\ on a covering space shortly.

\def\tmphm{(\tilde M,{\tilde q}_{ab})} Let $\tmphm$ be an arbitrary
homogeneous manifold, and $\tmphm/\Gam$ be its quotient. If we consider
a \diffeo\ $\phi:\;\tilde M\goes\tilde M$, then the following two
quotients are manifestly isometric;
\begin{equation}
        \tmphm/\Gam\simeq
        (\tilde M,\phi_*{\tilde q}_{ab})/\phi\inv\circ\Gam\circ\phi,
        \label{2-1}
\end{equation}
where $\phi_*$ is the ``pullback'' (i.e. the induced map) of $\phi$.
(Since both $\Gam$ and $\phi$ act on $\tilde M$, the composition of them
in the right hand side is well-defined.)  We can be benefited from this
relation in two (or more) practical contexts --- The first is just to
simplify the universal cover metric ${\tilde q}_{ab}$, while the second
is to deform the covering group $\Gam$ suitably.  Though we will utilize
the second in the next section, our main focus in this section is in the
first.  Another importance comes {\it when $\phi$ is an isometry} of the
universal cover $\tmphm$, i.e., $\phi_*{\tilde q}_{ab}={\tilde q}_{ab}$.
In this case, it holds
\begin{equation}
        \tmphm/\Gam\simeq
        (\tilde M,{\tilde q}_{ab})/\phi\inv\circ\Gam\circ\phi.
        \label{2-2}
\end{equation}
So, we can define an equivalence relation in the covering group by
\begin{equation}
        \Gam\sim\phi\inv\circ\Gam\circ\phi,
        \label{2-3}
\end{equation}
called {\it conjugation}.  With this, we can simplify the covering
group, even after we have exhausted \diffeos\ on $\tilde M$ to simplify
${\tilde q}_{ab}$.

Now, the subsequent procedure to get $\bar C$ is almost straightforward.
Introducing the equivalence relation by \diffeo\ in $U$, we obtain a
smooth set $\bar U$ of representative elements.  For convenience, we
often identify $\bar U$ with a universal covering spacetime with
$n=\dim\bar U$ smooth parameters, and denote these parameters and the
(parameteric) universal cover as \u\ and $\UCu$, respectively.

For a fixed $u\in\bar U$, we give all the possible \Gam's, and denote
the smooth set of the pairs, $(u,\Gam)$'s, as $C_u$. Then, we construct
the conjugacy class $\bar C_u$ of $C_u$ by \Isomf, i.e., equivalence
class defined by $(u,\Gam)\sim(u,a\circ\Gam\circ a\inv)$, where
$a\in\Isomf$.  We finally consider the smooth set $\bar C$ of spacetimes
obtained from $\bar C_u$ with varying $u$, i.e., $\bar C\equiv\{c|
c\in\bar C_u, u\in\bar U\}$.  We think of $\bar C$ as the natural
dynamical system we wanted, consisting of compact homogeneous universes.
Like $\bar U$, we identify $\bar C$ with a compact homogeneous universe
with $n+m$ smooth parameters, where $m=\dim\bar C_u$. We denote the $m$
parameters in $\bar C_u$ and the (parameteric) compact homogeneous
universe as, respectively, \g\ and $\UCu/\Gam_\g$.  The number of
dynamical degrees of freedom is now simply given by $\dim\bar C=\dim\bar
U+\dim\bar C_u=n+m$.  This completes our construction.

For example, when the extendible isometry group is given by a Bianchi
group, we can take, without loss of generality, the universal cover
metric as
\begin{equation}
        \d s^2=-\d t^2+\dh \mu\nu(t)\s\mu\s\nu,
        \label{2-5}
\end{equation}
where $\s\mu$ are the invariant 1-forms of the Bianchi group, and $\dh
\mu\nu$ is a nondegenerate symmetric $3\times3$ matrix function of $t$.
We substitute this into Einstein's equation, and then subtract the
degrees of freedom of \diffeos\ from the solution obtained.  We usually
do this last step with the {\it homogeneity preserving \diffeos\ (HPDs)}
\cite{AS}, which are defined by the condition that they induce
``rotations'' among the invariant 1-forms, i.e., for an HPD $\eta$, it
holds;
\begin{equation}
        \eta_*:\; \s\mu\goes f^\mu{}_\nu\s\nu,
        \label{2-6}
\end{equation}
where $f^\mu{}_\nu$ is a constant $3\times3$ matrix.  (Possible
$f^\mu{}_\nu$'s comprise a subgroup of $\GL3$.)  The final form of the
solution takes, in many cases, a diagonal form;
\begin{equation}
        \d s^2=-\d t^2+
        \dh11(t;\u)(\s1)^2+\dh22(t;\u)(\s2)^2+\dh33(t;\u)(\s3)^2.
        \label{2-7}
\end{equation}
This gives the set $\bar U$, or the metric $\udg ab[\u]$.  (The explicit
form of $\dh\mu\nu(t;\u)$ depends, of course, upon the given Bianchi
group.)  $\Gam_\g$ is obtained by exhausting possible conjugations with
respect to the extendible isometries, in the possible embeddings of the
fundamental group $\pi_1(M)$ into the extendible isometry group.

We note that the dynamical variables in configuration space are defined
with respect to the intrinsic geometry of the spatial section. Hence,
like that it was important to subtract the degrees of freedom of
diffeomorphisms and conjugations of the {\it spacetime} when
determining the number of dynamical degrees of freedom, it is important
to subtract the degrees of freedom of diffeomorphisms and conjugations
of the {\it space} to determine the values of the dynamical variables.
This process is basically the same as what has been presented for
spacetime, except for considering $\Isom\uMt$ instead of \Isomf.  In I,
we have presented parameterizations for almost all compact locally
homogeneous three-manifolds --- Each parameterization is like
\begin{equation}
  \label{2-4}
  \UCstcur/A_{\tei},
\end{equation}
where the standard universal cover \UCstcur\ is parameterized so as to
contain no degrees of freedom of diffeomorphisms on $\tilde M$, and the
covering group $A_{\tei}$ acting on \UCstcur\ is parameterized so as to
be free from the conjugations with respect to the (intrinsic) isometry
group $\Isom\tilde M\simeq\Isom\uMt$. As in II, we call the parameters
$(\cur,\tei)$ the dynamical variables.  Each compact locally homogeneous
spatial section $\UChu/\Gam_\g$ of the solution $\UCu/\Gam_\g$ is
isometric to $\UCstcur/A_{\tei}$ for some \cur\ and \tei.  The dynamical
variables $(\cur,\tei)$ can therefore be identified with functions of
time $\cur=\cur(t; \u)$, and $\tei=\tei(t; \u,\g)$, containing (\u,\g)
as constant parameters.  (\cur\ is determined only from the universal
cover metric $\udh ab[\u]$, so that \cur\ is independent of \g.) When
considering a Bianchi group as the extendible isometry group, \cur\ and
$\tei$ depend on $t$ and \u\ through the component functions
$\h\equiv(\dh11(t;\u),\dh22(t;\u),\dh33(t;\u))$ in Eq.\reff{2-7}, so
that we may write as $\cur=\cur(\h)$, and $\tei=\tei(\h,\g)$.  We
executed this procedure for some explicit models in II.

For convenience, we often reparametrize $\hst{[\cur]}$ or $A_{\tei}$,
and factor out $v$ to parameterize the volume.  In such a case, the
dynamical variables are \tcv, and they are functions of time containing
(\u,\g) as constant parameters.  We shall refer to the steps here of
finding the time-development of the dynamical variables as the {\it
  spacetime approach}, in distinction from the dynamical approach which
will be presented in the next section.

\section{Fixing covering maps and the natural Hamiltonian structures}
\label{sec3}

As is well known, general relativity possesses Hamiltonian formalism
(e.g. \cite{Wa}).  In this section, we consider how we can obtain, from
the full phase space, a reduced phase space for our dynamical system of
compact homogeneous universe. In our mathematical representation of a
compact homogeneous universe in the previous section, part of dynamical
degrees of freedom was contained in the covering map, which cannot be a
canonical dynamical variable in ordinary way.  In some cases, however,
we can move to another representation in which the covering map contains
no dynamical degrees of freedom and the universal cover metric carry all
the ones. In these cases we may obtain the natural reduced Hamiltonian.

We first consider a reparametrization of compact locally homogeneous
3-manifolds. Remember that our standard parameterization of compact
locally homogeneous 3-manifolds is like Eq.\reff{2-4}.  (For simplicity,
we do not factor out the total volume $v$ in this section.)  This
parameterization is of particular importance in that it is the most
natural one among those of containing no redundant parameters.  However,
once we find such a parameterization, we can reparametrize the
variations of the manifold appropriately, following Eq.\reff{2-1}. In
particular, we can fix the covering map, using a \diffeo\ 
$\phi_{\tei}:\;\Mtil\goes\Mtil$ such that
\begin{equation}
        A_{\tei}=\phi_{\tei}\circ A_0\circ\phi_{\tei}\inv,
        \label{3-1}
\end{equation}
where $A_0$ is the covering group for a set of fixed \Teich parameters
$\tei=\tei_0$.  With this \diffeo, we obtain parameterization
\begin{equation}
        \UCdycurtei/A_0,
        \label{par2}
\end{equation}
where
\begin{equation}
        \hdy{[\cur,\tei]}\equiv\phi_{\tei*}\hst{[\cur]}.
        \label{3-2}
\end{equation}
It is worth noting that $\phi_{\tei}$ is a \diffeo\ which maps a
fundamental region of the projection map $\pi_0$ of $A_0$ to a
fundamental region of $\pi_{\tei}$ of $A_{\tei}$.  We shall refer to
$\phi_{\tei}$ as a {\it \Teich \diffeo} (TD).

We may be able to see the metric \reff{3-2} as a ``dynamical'' metric.
That is, we expect there exists a spacetime metric $\gdy{(\cur,\tei)}$
whose spatial part is given by Eq.\reff{3-2}, where $\cur$ and $\tei$
are functions of time $t$.  If we start with such a ``dynamical''
spacetime metric with $\cur(t)$ and $\tei(t)$ being free functions of
$t$, then we would be able to find the time-development of them directly
from Einstein's equation. (Note that by definition, the quotient of the
solution obtained in this way is free from the degrees of freedom of
\diffeos.)  This may give another approach, which we call the {\it
  dynamical approach}, to obtain the time-development of the dynamical
variables $(\cur(t),\tei(t))$.  This is similar to the treatment
employed in (2+1)-gravity \cite{HNCa}.

We notice, however, that TDs are not unique, so that many possibilities
of different parameterizations of Eq.\reff{par2} exist. Moreover, even
for a given parameterization of Eq.\reff{par2}, there seem to exist many
possibilities of ways of taking the dynamical spacetime metric, i.e., of
taking the shift vector.  (The spatial dependence of the lapse function
results in an undesirable change of spatial foliation, but just
$t$-dependence of the lapse function simply results in a
reparametrization of $t$, which is not essential for our purpose.  We
therefore fix the lapse function to unity.)  However, the ambiguity of
the shift vector is not essential. What is essential is the choice of
TD.  To see this, suppose we are given the solution constructed
unambiguously in the spacetime approach for given \Isomf\ and
$\pi_1(M)$.

For an SCH spacetime to yield $\UChdycurtei/A_0$ as its spatial part,
this SCH spacetime need be of the form $\UCgdycurtei/A_0$ with $A_0$
being a discrete subgroup of the extendible isometry group.  Moreover,
this SCH spacetime must be isometric to the SCH spacetime constructed in
the spacetime approach, $\UCu/\Gam_\g$.  Hence, we must have the
following commutative diagram;
\begin{equation}
\begin{array}{ccccc}
  \UCx{[\u](t)}/\Gam_\g  & & \stackrel{\psi_\g}{\longleftarrow} & &
  \UCgdyx{(\cur(t),\tei(t))}/A_0 \\
  \downarrow             & &        & & \downarrow       \\
  \UChx{[t,\u]}/\Gam_\g &\leftarrow& \UCstx{[\cur(t)]}/A_{\tei(t)} &
             \stackrel{\phi_{\tei}}{\leftarrow} & 
             \UChdyx{[\cur(t),\tei(t)]}/A_0
\end{array}.
        \label{eq:commu}
\end{equation}
Here, the vertical arrows stand for restricting to spatial manifolds,
the horizontal arrows for mapping by \diffeos, and we have inserted
$t$-dependence explicitly.  [We are consistently using square brackets [
] to denote the parametric dependence of the arguments, while
using parentheses ( ) to denote the usual dependence of the arguments.]

Let $\psi_\g$ be the \diffeo\ $\psi_\g:\; \uMM\goes\uMM$ which connects
the two spacetime manifolds in Eq.\reff{eq:commu}. That is, $\psi_\g$ is
a \diffeo\ such that
\begin{equation}
        \Gam_{\g}=\psi_{\g}\circ A_0\circ\psi_{\g}\inv.
        \label{3-3}
\end{equation}
Also, $\psi_\g$ preserves each spatial section (i.e., preserves the
spatial foliation).  With $\psi_\g$, the two universal cover metrics
relate through
\begin{equation}
        \gdy{(\cur,\tei)}=\psi_{\g *} \udg ab[\u].
        \label{3-4}
\end{equation}
Note that the \diffeo\ which connects the left two spatial manifolds in
Eq.\reff{eq:commu} is uniquely specified. For example, when the
extendible isometry group is given by a Bianchi group, it was the
composition, $\eta\circ\iota:\; \tilde M\goes\uMt$, of an HPD $\eta$,
used to simplify the universal cover metric to the standard metric, and
an isometry $\iota$, corresponding to the conjugation. This, in turn,
implies that the TD $\phi_{\tei}$ and $\psi_\g$ are in one-to-one
correspondence;
\begin{equation}
        \psi_\g=\eta\circ\iota\circ\phi_{\tei}.
        \label{3-5}
\end{equation}
(Since $\psi_\g$ preserves the spatial foliation, $\psi_\g$ can be
naturally identified with spatial \diffeos. Conversely, since $\eta$,
$\iota$, and $\phi_{\tei}$ contain $t$ as parameter, they can be
naturally identified with \diffeos\ on the spacetime manifold.
Eq.\reff{3-5} is therefore well-defined on both $\uMt$ and $\uMM$. Note,
however, that $\iota$ is {\it not} in general an isometry when acting on
the spacetime $\UC$.)  Thus, we can focus on $\psi_\g$'s, instead of
thinking of TDs.  We may always find many varieties of foliation
preserving \diffeos\ $\psi_\g$'s which satisfy Eq.\reff{3-3}, so that
there still seem to exist many varieties of dynamical spacetime metrics.
Note, however, that for arbitrary choices of $\psi_\g$, the right hand
side of Eq.\reff{3-4} would depend freely on $t$, $\u$ and $\g$;
\begin{equation}
        \psi_{\g*}\udg ab[\u](t)=(\psi_{\g*}\udg ab)[\u,\g](t).
        \label{3-6}
\end{equation}
Since we cannot, in general, ``invert'' the functions $
\cur=\cur(t;\u)$, and $\tei=\tei(t;\u,\g) $, i.e., we cannot find
$t=t(\cur,\tei)$, $\u=\u(\cur,\tei)$, and $\g=\g(\cur,\tei)$, we would
not be able to have the right dynamical metric as a functional of $\cur(t)$
and $\tei(t)$;
\begin{equation}
        (\psi_{\g*}\udg ab)[\u,\g](t)\neq \gdy{(\cur(t),\tei(t))}.
        \label{3-8}
\end{equation}
Obviously, to have the right dynamical metric, the pullback of $\udg
ab[\u]$ by $\psi_\g$ must take a special form where $t$, $\u$, and $\g$
appears as algebraic combinations of $ \cur=\cur(t;\u)$ and
$\tei=\tei(t;\u,\g) $.  This shows the origin of the fact that an
arbitrary choice of TD in the dynamical approach leads us to only a bad
ansatz for the solution of the field equation, no matter how we choose
the shift vector.  Conversely, if we choose a right TD, the right
dynamical metric is unique due to a correspondence such as
Eq.\reff{3-5}.

In general it is not trivial to find right TDs. Even worse, in some
cases they may not exist. However, in special cases shown below the
existence of the right TDs is guaranteed and we can carry out the
dynamical approach.

Suppose the extendible isometry group is given by a Bianchi group $G$,
and consider a class of SCH spacetimes such that we can find the TDs in
the HPDs. We call the TDs implemented in (a subgroup of) the HPDs the
{\it homogeneity preserving \Teich \diffeos\ (HPTDs)}. The dynamical
metric (with zero shift vector) with the HPTDs becomes of the form
\reff{2-5}, which is expected to consistently describe the
time-development of the dynamical variables, since it is just an
ordinary, but non-diagonal generally, Bianchi type spacetime metric.
$\Isomf=G$ is also satisfied. (One might still care about that possible
momentum constraints would excessively reduce the number of dynamical
degrees of freedom, but this is not the case.  The momentum constraints
correspond to the inner automorphisms, Inn$(G)$, of the Bianchi group
$G$ \cite{AS}, which in turn correspond to the conjugations in the
quotient space. Since we have so defined the TDs that no conjugations
are contained in the quotient space, the dynamical metric will be free
from the momentum constraints.) Thus, the HPTDs will give a right
dynamical metric. This constitutes an essential ingredient of our
actual reduction of the Hamiltonian.

Finally, we can easily find a necessary condition to have the HPTDs,
which is given by
\begin{equation}
  \label{3-nec}
  \chi\geq\dim\bar C_u,
\end{equation}
where $\dim\bar C_u$ equals the number of the parameters $\g$,
and 
\begin{equation}
  \label{3-chi}
  \chi\equiv\dim{\rm Out}(G)=\dim{\rm HPDs}-\dim{\rm Inn}(G).
\end{equation}
Here, ${\rm Out}(G)$ is the outer automorphism group of $G$, and we have
used the fact that the HPDs comprise the automorphism group of $G$. The
condition \reff{3-nec} can be understood from the observation that the
$\psi_\g$ must be (time-independent) HPDs, modulo the gauge orbits
generated by the momentum constraints.

In the next section, we demonstrate the dynamical approach and show the
Hamiltonians by means of the HPTDs.

\section{Four examples}
\label{sec4}

\def\barf{{\bar f}^3{}_3}
\def\confhb{\paren{\frac{v}{(\a11)^2(\a22)^2}}^{2/3}}

We in this section apply our argument to the four vacuum models
investigated previously in II, namely the b/1, f1/1($n$), a1/1, and a2/1
models.  The definitions of the four compact homogeneous universes are
summarized in the table below, where the extendible isometry groups and
the fundamental groups of the spatial sections are given.  We have
abbreviated the Bianchi N group as ``BN''.  Explicit representations of
$\pi_1(M_{\rm b/1})$ and $\pi_1(M_{{\rm f1/1}(n)})$ are presented in,
respectively, Eqs.\eqnII{14} and \eqnII{37}.  (For simplicity, we in
this section refer to equations in I and II as \eqnI1, \eqnII1, etc..)

\medskip
\begin{center}
  {\small
\begin{tabular}{|c|c|c|c|}  \hline
        Model & \Isomf\ & Multiplication rule for \Isomf\ &
                                 $\pi_1$ \\ \hline\hline
        b/1 & BII & 
    $\vector{g^1}{g^2}{g^3}\vector{h^1}{h^2}{h^3}=
    \vector{g^1+h^1}{g^2+h^2}{g^3+h^3+g^1h^2} $ &
        $\pi_1(M_{\rm b/1})$ \\ \hline
        f1/1($n$) & B\sz &
    $ \vector{g^1}{g^2}{g^3}\vector{h^1}{h^2}{h^3}=
      \vector{g^1+e^{-g^3}h^1}{g^2+e^{g^3}h^2}{g^3+h^3} $ &
         $\pi_1(M_{\rm f1/1(n)})$ \\ \hline
        a1/1 & B\svz & 
    $ \vector{g^1}{g^2}{g^3}\vector{h^1}{h^2}{h^3}=
      \svector{\svector{g^1}{g^2}+R_{g^3}\svector{h^1}{h^2}}{g^3+h^3} $ &
        $\pi_1(T^3)$ \\ \hline
        a2/1 & BI &
    $ \vector{g^1}{g^2}{g^3}\vector{h^1}{h^2}{h^3}=
      \vector{g^1+h^1}{g^2+h^2}{g^3+h^3} $ &
         $\pi_1(T^3)$ \\ \hline
\end{tabular}
}
\end{center}
\medskip In this table, $R_{g^3}$ is the rotation matrix by angle $g^3$,
and the usual multiplication rule for matrix is understood there.  The
multiplication rules in this table will be applied without any notice in
the subsequent subsections.

We will see that the HPTDs are found, the dynamical metrics can be
written, and thereby the Hamiltonians are obtained, for the b/1,
f1/1($n$), and a2/1 models. For these models, $\chi=\dim\bar C_u$ holds,
so that the condition \reff{3-nec} is satisfied. The a1/1 model does not
satisfy this condition (but $0<\chi<\dim\bar C_u$). Nevertheless, we will
succeed to have a Hamiltonian also for this model, if we admit a
``degeneracy'' of the dynamical variables. (See Sec.\ref{a1}.)

All the arguments in the following subsections are basically parallel to
the first one, Sec \ref{b1}.

\subsection{The b/1 model}
\label{b1}

We begin with preparing the HPDs $\eta$ of Bianchi II, which are
obtained from the invariance under Eq.\reff{2-6} of the Maurer-Cartan
relation
\begin{equation}
        \d\s1=0,\quad \d\s2=0,\quad \d\s3=-\s1\^\s2.
        \label{b1-mc}
\end{equation}
The invariant 1-forms $\s i$ here of Bianchi II are defined by
\begin{equation}
        \s1\equiv\d x,\; \s2\equiv\d y,\; \s3\equiv\d z-x\d y,
        \label{b1-base}
\end{equation}
using a local coordinate basis $(x,y,z)$.  We find
\begin{equation}
        \f13=\f23=0,\quad \f33=\barf\equiv \f11\f22-\f12\f21,
        \label{b1-1}
\end{equation}
and so that
\begin{equation}
        \eta_* : \vector{\s1}{\s2}{\s3} \goes
        \matrix{\f11}{\f12}0{\f21}{\f22}0{\f31}{\f32}{\barf}
        \vector{\s1}{\s2}{\s3}.
        \label{b1-hpd*}
\end{equation}
Moreover, by integration we have
\begin{equation}
        \eta : \vector xyz\goes
        \vector{\f11x+\f12y}{\f21x+\f22y}
        {(1/2)(\f11\f21x^2+\f12\f22y^2)+\f12\f21xy+\barf z+\f31x+\f32y}.
        \label{b1-hpd}
\end{equation}
We have set the integral constants, corresponding to isometries, zero.
This is the general form of HPDs, up to isometry.  There are six free
parameters in it, which are constant with respect to the spatial
coordinates $(x,y,z)$, but possibly depend on time $t$, depending upon
the context.

Our standard universal cover metric, up to constant conformal factor, is
(cf. Eq.\eqnI{75})
\begin{equation}
        \d l^2=(\s1)^2+(\s2)^2+(\s3)^2,
        \label{hb-12}
\end{equation}
and the parameterization of the \Teich space is given by the following
generators acting on the standard universal cover (cf. Eq.\eqnI{129});
\begin{equation}
        A_{\tei}=\brace{ \vector{\a11}00,
        \vector{\a21}{\a22}0,
        \vector 00{\a11\a22} }.
        \label{b1-Atei}
\end{equation}
So, there are no universal cover parameters, and three \Teich
parameters;
\begin{equation}
        \cur=\emptyset,\quad \tei=(\a11,\a21,\a22).
\end{equation}
We take $\tei_0=(1,0,1)$, for later convenience.  The dynamical
variables are $\tei$ and the volume $v$, which has been omitted from the
standard universal cover metric basically for simplicity, but will come
into the dynamical metric finally.

We can choose the unit cube (Fig.1) in $0\leq x\leq1$, $0\leq y\leq1$
and $0\leq z\leq1$ as a fundamental region of $\pi_0$.  If we choose a
fundamental region of $\pi_\tau$ such that the projection onto the
$x$-$y$ plane is a parallelogram shown in Fig.2, then the HPTDs are
determined by the requirement that they comprise a subgroup of HPDs.  In
fact, for the $x$-$y$ plane, TD $\phi_\tau$ must be the linear
transformation
\begin{equation}
        \phi_{\tei}:\, \svector xy\goes \svector{\a11 x+\a21 y}{\a22 y},
        \label{hb-13}
\end{equation}
so that we have
\begin{equation}
        \phi_{{\tei}*}:\, \svector{\s1}{\s2}\goes
        \smatrix{\a11}{\a21}{0}{\a22}\svector{\s1}{\s2}.
        \label{hb-14}
\end{equation}
This in turn implies (cf. Eq.\reff{b1-hpd*})
\begin{equation}
        \phi_{{\tei}*}:\, \vector{\s1}{\s2}{\s3}\goes
        \matrix{\a11}{\a21}{0}{0}{\a22}{0}{0}{0}{\a11\a22}
        \vector{\s1}{\s2}{\s3}.
        \label{hb-15}
\end{equation}
It is easy to observe that the TDs of Eq.\reff{hb-15} certainly comprise
a subgroup of HPDs.  Comparing with Eq.\reff{b1-hpd}, we finally have
\begin{equation}
        \phi_{\tei}:\, \vector xyz\goes
        \vector{\a11 x+\a21 y}{\a22 y}{\rcp2\a21\a22 y^2+\a11\a22 z}.
        \label{hb-16}
\end{equation}
The image of the unit cube by this ``HPTD'' must be a fundamental region
of $\pi_\tau$ (Fig.1), i.e., must satisfy Eq.\reff{3-1}.  In fact, for
the generators of $A_0$
\begin{equation}
  \label{G0}
  A_0=\brace{\vector100,\vector010,\vector001},
\end{equation}
we can calculate
\begin{equation}
  \label{Gtau}
  \phi_{\tei}\circ A_0\circ\phi_{\tei}\inv=
  \brace{\vector{\a11}00,\vector{\a21}{\a22}{\rcp2\a21\a22},
    \vector00{\a11\a22}},
\end{equation}
which certainly coincides with $A_\tau$, up to conjugation.  (The third
component of the second generator of Eq.\reff{Gtau} differs from that of
Eq.\reff{b1-Atei}, but this is not essential for our purpose, since we
can make it zero by using an isometry $\iota$, i.e.,
$\phi_{\tei}\goes\phi_{\tei}\circ\iota$ in Eq.\reff{Gtau}. This change
does not affect the form of the dynamical metric.)  Thus,
Eq.\reff{hb-16} is the right HPTDs.

The induced metric of Eq.\reff{hb-12} by $\phi_{\tei}$ is obtained from
the direct substitution of Eq.\reff{hb-15}.  Normalizing the induced
metric to give $v^2$ as determinant, (and attaching ``$-\d t^2$'',) we
obtain the dynamical metric
\begin{equation}
        \d s^2=-\d t^2+\H\alpha\beta\s\alpha\s\beta,
        \label{hb-4}
\end{equation}
where
\begin{eqnarray}
        \H11\wa \confhb (\a11)^2, \nonumber \\
        \H22\wa \confhb ((\a21)^2+(\a22)^2), \nonumber \\
        \H12\wa \confhb \a11\a21, \nonumber \\
        \H33\wa \confhb (\a11)^2(\a22)^2.
        \label{hb-7}
\end{eqnarray}
The inverse is also useful;
\begin{equation}
        \a11=\H33\sqrt{\frac{\H11}{H}},\,
        \a21=\frac{\H33\H12}{\sqrt{H\H11}},\,
        \a22=\sqrt{\frac{\H33}{\H11}},\,v=\sqrt H,
        \label{hb-8}
\end{equation}
where $H\equiv\det(\H\alpha\beta)=(\H11\H22-(\H12)^2)\H33$.  Now, we can
think of the set of metric components $(\H11,\H22,\H12,\H33)$, which are
functions of $t$, as an alternative set of the dynamical variables.  In
such a case, the geometrical meaning of them should be understood with
respect to Eq.\reff{hb-7}.

We can also find the \diffeo\ $\psi_\g$ on $\uMM$ defined by
Eq.\reff{3-3}, which is given by
\begin{equation}
        \psi_{\g*}:\; \vector{\s1}{\s2}{\s3}\goes 
        \matrix{\dug11}{\dug21}0{\dug12}{\dug22}000{\bar g_3{}^3}
        \vector{\s1}{\s2}{\s3},
        \label{psig*}
\end{equation}
or
\begin{equation}
        \psi_{\g}:\; \vector xyz\goes 
        \vector{\dug11x+\dug21y}{\dug12x+\dug22y}
        {(1/2)(\dug11\dug12x^2+\dug21\dug22y^2)+\dug12\dug21xy+
          \bar g_3{}^3z},
        \label{psig}
\end{equation}
where $\bar g_3{}^3\equiv \dug11\dug22-\dug12\dug21$.  In fact, we can
easily calculate
\begin{equation}
        \psi_{\g}\circ A_0\circ\psi_{\g}\inv=
        \brace{\vector{\dug11}{\dug12}{\rcp2\dug11\dug12},
        \vector{\dug21}{\dug22}{\rcp2\dug21\dug22},
        \vector{0}{0}{\bar g^3{}_3}},
        \label{b1-x1}
\end{equation}
which coincides with $\Gam_\g$ given in Eq.\eqnII{18}, up to unessential
conjugation, again.  (Note that Eq.\reff{psig*} corresponds to a case of
$\f31=\f32=0$ for the general HPDs \reff{b1-hpd*}. The spacetime metric
with $\f31$ and $\f32$ as constant parameters was possible as a solution
of Einstein's equation, but the space spanned by $\f31$ and $\f32$ in
the space of solutions corresponds to the gauge orbits generated by the
momentum constraints, so that they cannot be true dynamical degrees of
freedom.)

If we make the dynamical metric following Eq.\reff{3-4} (with the
universal cover metric \reff{2-7}), and substitute into Eq.\reff{hb-8},
we obtain the time development in terms of $\h$ of the dynamical
variables, previously obtained in II (see Eqs.\eqnII{31} and
\eqnII{32}).  This guarantees the rightness of the dynamical metric
\reff{hb-4}.

Finally, our phase space $P$ is spanned by
$(\a11,\a21,\a22,v,\p11,\p21,\p22,p_v)$, where $\p ij$ and $p_v$ are the
conjugate momenta of $\a ij$ and $v$, respectively.  The Hamiltonian on
$P$ is naturally obtained from the dynamical metric \reff{hb-4} and the
usual Einstein-Hilbert action as
\begin{eqnarray}
        {\cal H}\wa\rcp{2v}\left\{ (\a11\a22v)^{4/3}+(\a11)^2(\p11)^2
        +((\a21)^2+(\a22)^2)(\p21)^2+(\a22)^2(\p22)^2 \right. \nonumber \\
        && \left. +2\a11\a21\p11\p21+\a11\a22\p11\p22+\a21\a22\p21\p22
        \right\}
        -{3\over8}v(p_v)^2.
        \label{b1-H}
\end{eqnarray}
The Hamiltonian constraint, given by ${\cal H}\approx0$, reduces the
dynamical degrees of freedom from $\dim P=8$ to six, which agrees with
the count given in II, i.e., two for \u, and four for
$\g=\brace{\dug11,\dug12,\dug21,\dug22}$.

\subsection{The f1/1$(n)$ model}

As in Bianchi II, the HPDs $\eta$ of Bianchi \sz\ are obtained from the
invariance under Eq.\reff{2-6} of the Maurer-Cartan relation
\begin{equation}
        \d\s1=\s2\^\s3,\quad \d\s2=-\s3\^\s1,\quad \d\s3=0,
        \label{b6-mc}
\end{equation}
with respect to the invariant 1-forms $\s i$ of Bianchi \sz, given by
\begin{equation}
        \s1\equiv\rcp{\sqrt2}(e^z\d x+e^{-z}\d y),\,
        \s2\equiv\rcp{\sqrt2}(-e^z\d x+e^{-z}\d y),\,
        \s3\equiv\d z.
        \label{f1-base}
\end{equation}
We find
\begin{equation}
        \eta_*: \vector{\s1}{\s2}{\s3}\goes
        \matrix{\f11}{\f12}{\f13}{\pm\f12}{\pm\f11}{\f23}{0}{0}{\pm1}
        \vector{\s1}{\s2}{\s3}.
        \label{f1-hpd*}
\end{equation}
We need only the identity component of the HPDs, which correspond to the
plus signs in Eq.\reff{f1-hpd*}.  Integrating $\eta_*$ for the plus
signs, we have
\begin{equation}
        \eta : \vector xyz\goes
        \vector{(\f11-\f12)x-\rcp{\sqrt2}(\f13-\f23)e^{-z}}
        {(\f11+\f12)y+\rcp{\sqrt2}(\f13+\f23)e^{z}}
        {z}.
        \label{f1-hpd}
\end{equation}
We have again set the integral constants zero.

Our standard universal cover metric is (cf. Eq.\eqnI{80})
\begin{equation}
        \d l^2=e^{2\lambda}(\s1)^2+e^{-2\lambda}(\s2)^2+(\s3)^2,
        \label{f1-std}
\end{equation}
and our parameterization of the \Teich space is given by the following
generators acting on the standard universal cover (cf. Eqs.\eqnI{156}
and \eqnI{157});
\begin{equation}
        A_{\tei}=\brace{ \vector{\alpha u_1}{\alpha u_2}0,
        \vector{\alpha v_1}{\alpha v_2}0,
        \vector 00{c_3} }, \quad{\rm for}\quad n>2,
        \label{f1-Atei1}
\end{equation}
and
\begin{equation}
        A_{\tei}=\brace{ \vector{\alpha u_1}{\alpha u_2}0,
        \vector{\alpha v_1}{\alpha v_2}0,
        h\circ\vector 00{c_3} }, \quad{\rm for}\quad n<-2,
        \label{f1-Atei2}
\end{equation}
where $e^{c_3}\equiv\abs{n+\sqrt{n^2-4}}/2$,
\def\sptmp{\sqrt{\abs{n+\sqrt{n^2-4}}\over2}}
\def\smtmp{\sqrt{\abs{n-\sqrt{n^2-4}}\over2}}
\begin{equation}
        \svector{u_1}{v_1}\equiv
        \rcp{\sqrt{\abs n}}\svector{\sptmp}{\smtmp},\;
        \svector{u_2}{v_2}\equiv
        \rcp{\sqrt{\abs n}}\svector{\smtmp}{\sptmp},
        \label{f1-uv}
\end{equation}
and $h$ is defined by $h :\; (x,y,z)\goes (-x,-y,z)$.  So, we have
\begin{equation}
        \cur=\brace{\lambda},\quad \tei=\brace{\alpha}.
        \label{f1-1}
\end{equation}
Adding the volume $v$, we have three dynamical variables
$(\cur,\tei,v)$.

It is quite easy to find the (HP)TDs, if noting the fact that the
compact manifold f1/1$(n)$ is a torus bundle over the circle.  Since the
\Teich parameter $\alpha$ is the size of the torus fiber, laid in
$x$-$y$ plane, relative to the circle, laid in z-axis, the HPTDs are
supposed to be
\begin{equation}
        \phi_{\tei}:\; (x,y,z)\goes (\alpha x,\alpha y, z).
        \label{f1-hptd}
\end{equation}
This is apparently an element of the HPDs \reff{f1-hpd}, i.e.,
corresponding to the case of $\f11=\alpha,\f12=\f13=\f23=0$.  We can
even make sure the relation \reff{3-1} if letting $\tei_0=\brace{1}$.
The \diffeos\ \reff{f1-hptd} are therefore the HPTDs.

As in the b/1 case, normalizing the induced metric of Eq.\reff{f1-std}
to give $v$ as its volume element, we obtain the dynamical metric
\begin{equation}
        \d s^2=-\d t^2+\H11(\s1)^2+\H22(\s2)^2+\H33(\s3)^2,
        \label{f1-dyn}
\end{equation}
where \def\conff{\paren{v^2\alpha^{-4}}^{1/3}}
\begin{eqnarray}
        \H11\wa\conff e^{2\lambda}\alpha^2, \nonumber \\
        \H22\wa\conff e^{-2\lambda}\alpha^2, \nonumber \\
        \H33\wa\conff. 
        \label{f1-dyncomp}
\end{eqnarray}
The inverse is also useful;
\begin{equation}
        \alpha=\frac{(\H11\H22)^{1/4}}{\sqrt{\H33}},\;
        \lambda=\rcp4\ln{\H11\over\H22},\; v=\sqrt{\H11\H22\H33}.
        \label{f1-dyncompinv}
\end{equation}

We can easily check that the $\psi_\g$ defined by Eq.\reff{3-3} with
$\g=\brace{\alpha_0}$ is given by
\begin{equation}
        \psi_{\g}:\; (x,y,z)\goes (\alpha_0x,\alpha_0y,z).
        \label{f1-psig}
\end{equation}
In the same manner as the b/1 case, we obtain the time development of
the dynamical variables,
\begin{equation}
        \alpha={(\dh11\dh22)^{1/4}\over\sqrt{\dh33}}\alpha_0,\quad
        \lambda=\rcp4\ln{\dh11\over\dh22},\quad
        v=\sqrt{\dh11\dh22\dh33}(\alpha_0)^2,
        \label{f1-timed}
\end{equation}
which coincides with Eq.\eqnII{46}.  Again, this guarantees the
rightness of the dynamical metric \reff{hb-4}.  (There are factor or
power errors in the last paragraph of Sec.IV B of II, for the values of
$\lambda$ and $v$.)

Our phase space $P$ is spanned by
$(\lambda,\alpha,v,p_\lambda,p_\alpha,p_v)$, where $p_\lambda$,
$p_\alpha$ and $p_v$ are the conjugate momenta of $\lambda$, $\alpha$
and $v$, respectively.  The Hamiltonian on $P$ is naturally obtained
from the dynamical metric \reff{f1-dyn} as
\begin{equation}
        {\cal H}=\rcp{2v}\left\{ 4\cosh^22\lambda\, (\alpha v)^{4/3}+
        \rcp4 p_\lambda^2+\frac34\alpha^2p_\alpha^2 \right\}
        -{3\over8}v(p_v)^2.
        \label{f1-H}
\end{equation}
The Hamiltonian constraint ${\cal H}\approx0$ reduces the dynamical
degrees of freedom from $\dim P=6$ to four, which agrees with the count
given in II, i.e., three for \u, and one for $\g=\brace{\alpha_0}$.


\subsection{The a1/1 model}
\label{a1}

The pullback of HPDs for Bianchi \svz\ are found to be
\begin{equation}
        \eta_*: \vector{\s1}{\s2}{\s3}\goes
        \matrix{\f11}{\f12}{\f13}{\mp\f12}{\pm\f11}{\f23}{0}{0}{\pm1}
        \vector{\s1}{\s2}{\s3},
        \label{a1-hpd*}
\end{equation}
for the invariant 1-forms defined by
\begin{equation}
        \s1=\cos z\d x+\sin z\d y,\,
        \s2=-\sin z\d x+\cos z\d y,\,
        \s3=\d z.
        \label{a1-ibase}
\end{equation}
We need only the identity component of the HPDs, which correspond to the
upper set of signs in Eq.\reff{a1-hpd*}.  Integrating $\eta_*$ for the
identity component, and setting the integral constants zero, we obtain
the following HPDs;
\begin{equation}
        \eta:\; \vector xyz\goes
        \vector{\f11x+\f12y+\f13\sin z+\f23\cos z}
        {-\f12x+\f11y-\f13\cos z+\f23\sin z}{z}.
        \label{a1-hpd}
\end{equation}

Our standard universal cover metric is (cf. Eq.\eqnI{72})
\begin{equation}
        \d l^2=e^{2\lambda}(\s1)^2+e^{-2\lambda}(\s2)^2+(\s3)^2,
        \label{a1-std}
\end{equation}
and our parameterization of the \Teich space is given by the following
generators acting on the standard universal cover (cf. Eq.\eqnI{100});
\begin{equation}
        A_{\tei}=\brace{ \vector{\a11}0{2l\pi},
        \vector{\a21}{\a22}{2m\pi},
        \vector{\a31}{\a32}{2n\pi} },
        \label{a1-Atei}
\end{equation}
where $l$, $m$, and $n$ are integers.  (The choice of Eq.\eqnI{101} will
give the same dynamical metric as that of Eq.\eqnI{100}, since they
differ by just a discrete element.  So, we focus on the case of
Eq.\reff{a1-Atei}.)

Note that we have {\it seven} variables $(\cur,\tei,v)$, where
\begin{equation}
        \cur=\brace{\lambda},\quad \tei=\brace{\a11,\a21,\a22,\a31,\a32},
        \label{a1-curtei}
\end{equation}
and $v$ is the volume.  We know, however, that all the \Teich parameters
\tei\ develop in time {\it in the same manner} (see Eq.\eqnII{54});
\begin{equation}
  \label{a1-x1}
  \a ij=\frac{(\dh11\dh22)^{1/4}}{\sqrt{\dh33}}\dug ij,\, 
  (i,j)=(1,1),(2,1),(2,2),(3,1),{\rm and}\, (3,2),
\end{equation}
where $\g\equiv\brace{\dug ij}$ are the constant parameters in the
$\Gam_\g$ (cf. Eq.\eqnII{50}).  This may suggest that dynamically, and
at least classically, the five \Teich parameters degenerate and we
should think of only one of them as a true dynamical variable, though
geometrically all of them reflect true degrees of freedom of smooth
deformations of the compact manifold a1/1.  We in this subsection take
this standpoint, since only by doing so we can succeed to have the
dynamical spacetime metric. (See below.)

Now, our dynamical variables are $(\cur,\tei',v)$, where
$\tei'=\brace{a}$ is one of the elements in \tei.  We also define
$\g'\equiv\brace{g}$, which is one of \g, corresponding to $a$.  The
HPTDs, the $\psi_{\g'}$, and the dynamical spacetime metric are most
similar to those of the f1/1$(n)$ model.  The HPTDs are given by
\begin{equation}
        \phi_{\tei'}:\; (x,y,z)\goes(a x,a y,z),
        \label{a1-hptd}
\end{equation}
which is a case of $\f12=\f13=\f23=0$ of Eq.\reff{a1-hpd}.  The
$\psi_{\g'}$ is given by
\begin{equation}
        \psi_{\g'}:\; (x,y,z)\goes(g x,g y,z).
        \label{a1-psig}
\end{equation}
The dynamical spacetime metric is given by
\begin{equation}
        \d s^2=-\d t^2+\H11(\s1)^2+\H22(\s2)^2+\H22(\s2)^2,
        \label{a1-dyn}
\end{equation}
where \def\confa{\paren{v^2a^{-4}}^{1/3}}
\begin{eqnarray}
        \H11\wa\confa e^{2\lambda}a^2, \nonumber \\
        \H22\wa\confa e^{-2\lambda}a^2, \nonumber \\
        \H33\wa\confa. 
        \label{a1-dyncomp}
\end{eqnarray}
Then, as in the f1/1$(n)$ case, we should have
\begin{equation}
  \label{a1-x2}
  a=\frac{(\dh11\dh22)^{1/4}}{\sqrt{\dh33}}g,
\end{equation}
which reproduces (a representative of) Eq.\reff{a1-x1}.  In this sense,
metric \reff{a1-dyn} is the right dynamical metric.  This diagonal type
of metric was, in fact, the only possibility of the dynamical metric
made from the HPTDs, since if we wrote the general Bianchi \svz\ 
spacetime metric, the off-diagonal three components would have become
nondynamical, due to the three momentum constraints \cite{AS}.

Our phase space $P$ is spanned by $(\lambda,a,v,p_\lambda,p_a,p_v)$,
where $p_\lambda$, $p_a$ and $p_v$ are the conjugate momenta of
$\lambda$, $a$ and $v$, respectively.  The Hamiltonian on $P$ is
naturally obtained from the dynamical metric \reff{a1-dyn} as
\begin{equation}
        {\cal H}=\rcp{2v}\left\{ 4\sinh^22\lambda\, (a v)^{4/3}+
        \rcp4 p_\lambda^2+\frac34a^2p_a^2 \right\}
        -{3\over8}v(p_v)^2.
        \label{a1-H}
\end{equation}
The Hamiltonian constraint ${\cal H}\approx0$ reduces the dynamical
degrees of freedom from $\dim P=6$ to four, which does {\it not}, of
course, agree with the count given in II, because variable $a$ is a
representative of the five \Teich parameters $\a ij$.

\subsection{The a2/1 (flat torus) model}
\def\tmpa{{\bf a}}

Both the HPDs for Bianchi I and the pullback of them are the general
linear transformations;
\begin{equation}
        \eta:\; x^\mu\goes\sum_\nu\f\mu\nu x^\nu;\quad 
        \eta_*:\; \s\mu\goes\sum_\nu\f\mu\nu\s\nu,
        \label{a2-hpd}
\end{equation}
where $\det(\f\mu\nu)\neq0$, $(x^1,x^2,x^3)\equiv (x,y,z)$, and
$(\s1,\s2,\s3)\equiv(\d x,\d y,\d z)$.

Our standard universal cover is simply (cf. Eq.\eqnI{66})
\begin{equation}
        \d l^2=(\s1)^2+(\s2)^2+(\s3)^2,
        \label{a2-std}
\end{equation}
and our parameterization of the \Teich space is given by the following
generators acting on the standard universal cover;
\begin{equation}
        A_{\tei}=\brace{ \vector{\a11}00,
        \vector{\a21}{\a22}{0},
        \vector{\a31}{\a32}{\a33} }.
        \label{a2-Atei}
\end{equation}
That is, we have no curvature parameter and six \Teich parameters;
\begin{equation}
        \cur=\emptyset,\quad \tei=\brace{\a11,\a21,\a22,\a31,\a32,\a33}.
        \label{a2-curtei}
\end{equation}
We choose $\tei_0=\brace{1,0,1,0,0,1}$, for later convenience.  In the
parameterization \reff{a2-Atei}, the dynamical variables are only the
\Teich parameters \tei; the freedom of volume variations are contained
in them.  This is due to the fact that a conformal transformation on the
standard universal cover with constant conformal factor is just the
pullback of a \diffeo\ on the cover. This occurs only in the case of
Bianchi I.  If one wants to factor out the volume $v$ from the dynamical
variables, one can do so by normalizing \tei\ to give the unit volume
for the quotient, but we don't as in II.

Let $\tmpa=\vector{a^1}{a^2}{a^3},\x=\vector xyz\in {\rm BI}$.  Then,
\begin{eqnarray}
        \eta\circ\tmpa\circ\eta\inv\circ\x\wa
        \eta\circ\tmpa\circ(f\inv\x) \nonumber \\
        \wa \eta\circ (\tmpa+f\inv\x) \nonumber \\
        \wa f\tmpa+\x,
        \label{a2-x1}
\end{eqnarray}
where $f\equiv(\f\mu\nu)$ is the matrix composed of $\f\mu\nu$, and the
usual multiplication rule for matrix is understood whenever ``$\circ$''
does not appear.  Thus, the induced element of $\tmpa$ by $\eta$ is the
linear transformation of $\tmpa$;
\begin{equation}
        \eta\circ\tmpa\circ\eta\inv=f\tmpa.
        \label{a2-rl}
\end{equation}
The TDs must satisfy Eq.\reff{3-1}, which is now equivalent to
\begin{equation}
        fM_0=M_{\tei},
        \label{a2-3-1}
\end{equation}
where $M_{\tei}$ is the matrix composed of the generators of $A_{\tei}$;
\begin{equation}
        M_{\tei}\equiv\matrix{\a11}{\a21}{\a31}0{\a22}{\a32}00{\a33}.
        \label{a2-Mtei}
\end{equation}
Since $M_0$ is the identity matrix, this implies
\begin{equation}
        f=M_{\tei}.
        \label{a2-f}
\end{equation}
We have found the TDs in the HPDs, so this is the HPTDs.  The pullback
of the standard metric \reff{a2-std} by these HPTDs immediately gives
(the spatial part of) the dynamical metric; we have
\begin{equation}
        \d s^2=-\d t^2+\H\alpha\beta\s\alpha\s\beta,
        \label{a2-dyn}
\end{equation}
where
\begin{equation}
        (\H\alpha\beta)=
        (\sum_\gamma M_{\tei\gamma\alpha}M_{\tei\gamma\beta})=
    \matrix{(\a11)^2}{\a11\a21}{\a11\a31}
                {}{(\a21)^2+(\a22)^2}{\a21\a31+\a22\a32}
                {(sym.)}{}{(\a31)^2+(\a32)^2+(\a33)^2}.
        \label{a2-dyncomp}
\end{equation}

Now, the metric components $(\H11,\H12,\H13,\H22,\H23,\H33)$ are an
alternative set of the dynamical variables \tei.  We can use the general
Bianchi I spacetime metric \reff{a2-dyn} as the universal cover metric
of the flat torus (a2/1) universe.  In such a case, the geometrical
interpretation of the components should be understood with respect to
Eq.\reff{a2-dyncomp}.

The $\psi_\g$ defined by Eq.\reff{3-3} is found to be
\begin{equation}
        \psi_{\g*}:\; \s\mu\goes\sum_\nu\dug\nu\mu\s\nu,
        \label{a2-psig*}
\end{equation}
or
\begin{equation}
        \psi_{\g}:\; x^\mu\goes\sum_\nu\dug\nu\mu x^\nu.
        \label{a2-psig}
\end{equation}
In fact, we can easily check that Eq.\reff{3-3} holds for the $\Gam_\g$
given by (cf. Eq.\eqnII{58})
\begin{equation}
        \Gam_\g=\brace{\vector{\dug11}{\dug12}{\dug13},
        \vector{\dug21}{\dug22}{\dug23},
        \vector{\dug31}{\dug32}{\dug33}}.
        \label{a2-gamg}
\end{equation}
If we make the dynamical metric following Eq.\reff{3-4} (with the
universal cover metric \reff{2-7}), the spatial components
$\H\alpha\beta$ are found to be
$\H\alpha\beta=\sum_{\mu,\nu}\dh\mu\nu\dug\alpha\mu\dug\beta\nu$. Here,
$\dh\mu\nu=\diag{t^{2p_1},t^{2p_2},t^{2p_3}}$, and $p_\alpha$'s are
constant parameters satisfying $p_1+p_2+p_3=1=(p_1)^2+(p_2)^2+(p_3)^2$,
for vacuum.  Comparison with Eq.\reff{a2-dyncomp} will give the same
time development of the \Teich parameters as Eq.\eqnII{65}, which fact
justifies the use of our dynamical metric.

Finally, our phase space $P$ is spanned by the six $\a ij$'s and the
corresponding conjugate momenta $\p ij$'s.  Though the present
parameterization may not be convenient for practical purpose, we, for
completeness, present the Hamiltonian on $P$, which is naturally
obtained from the dynamical metric \reff{a2-dyn} with
Eq.\reff{a2-dyncomp};
\begin{eqnarray}
        {\cal H}=\rcp{8\a11\a22\a33} && \bigg\{
        \paren{\a11\p11+\a21\p21-\a22\p22+\a31\p31-\a32\p32-\a33\p33}^2
                 \nonumber \\
        &&
        +4 \paren{\a22\p21+\a32\p31}^2
        +4 (\a33)^2\paren{(\p31)^2+(\p32)^2}
                \nonumber \\
        &&  
                -4 \a33\p33\paren{\a22\p22+\a32\p32}
        \bigg\}.
        \label{a2-H}
\end{eqnarray}
The Hamiltonian constraint ${\cal H}\approx0$ reduces the dynamical
degrees of freedom from $\dim P=12$ to ten, which agrees with the count
given in II, i.e., one for \u\ (the independent Kasner parameter), and
nine for $\g=\brace{\dug\mu\nu}$ (the constant parameters needed for the
compactification).

\section{Concluding remarks}
\label{lastsec}

We have defined a class of \diffeos, called \Teich \diffeos (TDs), which
induce the \Teich deformations of the quotient, in conjunction with a
fixed covering group. A subclass of the TDs, called the HPTDs, has also
been defined. They match the dynamical evolution of the \Teich
parameters. We have obtained the Hamiltonians for the space $P=T_*F$, by
restricting the spacetime metric to the dynamical metric, i.e., to the
metric induced by the HPTDs.

The Hamiltonian structure thus found is the same as the one which would
have obtained from the general Bianchi metric which is free from the
momentum constraints. However \cite{AS}, the dynamical degrees of
freedom do not have clear meaning in the context of the usual {\it open}
model. In our {\it compact} model, in contrast to this, all the
dynamical variables, and thus all the dynamical degrees of freedom, have
explicit geometrical meaning, since our approach is ``constructive''.
One can, nevertheless, also regard our results as giving geometrical
interpretation of the known Hamiltonian for a Bianchi metric.

Our approach will be applicable to larger classes of spacetimes, e.g.,
spacetimes of which space is {\it non} compact locally homogeneous but
has nontrivial fundamental group, and spacetimes of which space
possesses less than three local Killing vectors. One in the former case
has less \Teich parameters than the compact case. Whether a consistent
Hamiltonian is obtained or not will, however, depend upon each model.

We end this final section with some remarks on the compact models on
Bianchi VIII, i.e., compact models such that (the identity component of)
the extendible isometry group is given by the Bianchi VIII group, i.e.,
$\Esom_0\uMt={\rm BVIII}$.  We simply call these models the ``d/$*$''
models, where, as usual, d stands for the type of the universal cover
and $*$ is a characterization of the compact quotient. (See I, for
explicit description of ``$*$''.) For definiteness, we show the standard
universal cover metric, up to constant conformal factor, of the d/$*$
models; it is given by
\begin{equation}
  \label{d-std}
  \d l^2=e^{2(\lambda_+/\sqrt3+\lambda_-)}(\s1)^2+
  e^{2(\lambda_-/\sqrt3-\lambda_-)}(\s2)^2+
  e^{-(4/\sqrt3)\lambda_+}(\s3)^2,
\end{equation}
where $\lambda_+$ and $\lambda_-$ are constants, and the invariant
1-forms $\s\mu$ are
\begin{equation}
  \label{d-inv}
  \s1=\rcp y (\sin z\d x-\cos z\d y),\quad
  \s2=\rcp y (\cos z\d x+\sin z\d y),\quad
  \s3=\rcp y (\d x+y\d z).
\end{equation}
(The present choice of the invariant 1-forms differs from the one listed
in I. This is, of course, a matter of preference. But it was an
erroneous claim of I that \cite{HK} to admit a compact quotient, the
universal cover should admit four dimensional isometry group and
therefore the $\lambda_-$ should vanish. Actually, there exist many
cases such that we can embed the fundamental group $\pi_1({\rm d}/*)$
into BVIII itself {\it with} the discrete but generic isometries $j:
z\goes z+2n\pi$, where $n$ is an integer. The coefficients of the
invariant 1-forms (i.e., the curvature parameters) in Tables I and II of
Paper I should, accordingly, read as Eq.\reff{d-std}, and the ``Degrees
of freedom of the universal cover U'' in Table III of the same paper
should read ``2''.)

We know that the vacuum solution of the conventional Bianchi VIII model
is diagonalizable (e.g. Ref.\cite{RS}), which means that the universal
cover metric $\udg ab[\u]$ of the spacetime approach is of the form of
Eq.\reff{2-7}. So, its spatial part $\udh ab[\u]$ is also diagonal. On
the other hand, the standard metric $\hst[\cur]=e^{2\alpha}(\d
l^2)_{ab}$, where $\d l^2$ is given by Eq.\reff{d-std} and $e^{2\alpha}$
is a conformal factor, is of the diagonal form of which three components
can be freely specified. So, the spatial metric $\udh ab[\u]$ can be
directly identified with the standard metric, i.e., $\hst[\cur]=\udh
ab[\u]$.  Moreover, since it holds $\Isomf=\Isom\uMt$, we do not need
take conjugations of $\Gam_\g$ to have it coincide with $A_{\tei}$.
Therefore, we have $\Gam_\g=A_{\tei}$, implying
\begin{equation}
  \label{d-tei}
  \tei(t)=\g={\rm constants}.
\end{equation}
Namely, the \Teich parameters of the d/$*$ models do not develop in
time. In this sense, the \Teich parameters of the d/$*$ models may be
regarded as nondynamical, though we defined any \Teich parameter as a
dynamical variable. If on this standpoint, the dynamical variables are
only the curvature parameters $(\lambda_+,\lambda_-,\alpha)$, or
equivalently the three diagonal components of the Bianchi VIII metric,
so that the reduced Hamiltonian with its geometrical interpretation is
trivial. In fact, since $\chi=0$ holds, the HPDs for Bianchi VIII have
no freedom to store the TDs.

\medskip

{\it Note after the completion of this work:} Recently, Kodama \cite{Ko}
has studied the canonical structure for SCH spacetimes by a different
approach, the \diffeo-invariant phase space approach, giving consistent
Hamiltonians with ours.

\ack

\eject

\section*{Figure captions}

\noindent{\bf FIGURE 1:} The unit cube (a fundamental region of
    $\pi_0$) and its image (a fundamental region of $\pi_\tau$) by an
    HPTD.

\noindent{\bf FIGURE 2:} The parallelogram shows the
    projection of a fundamental region of $\pi_\tau$ on the $x$-$y$
    plane.

\eject

\begin{figure}[t]
\label{fig:1}
  \begin{center}
    \leavevmode
\epsfysize=18cm
\epsfbox{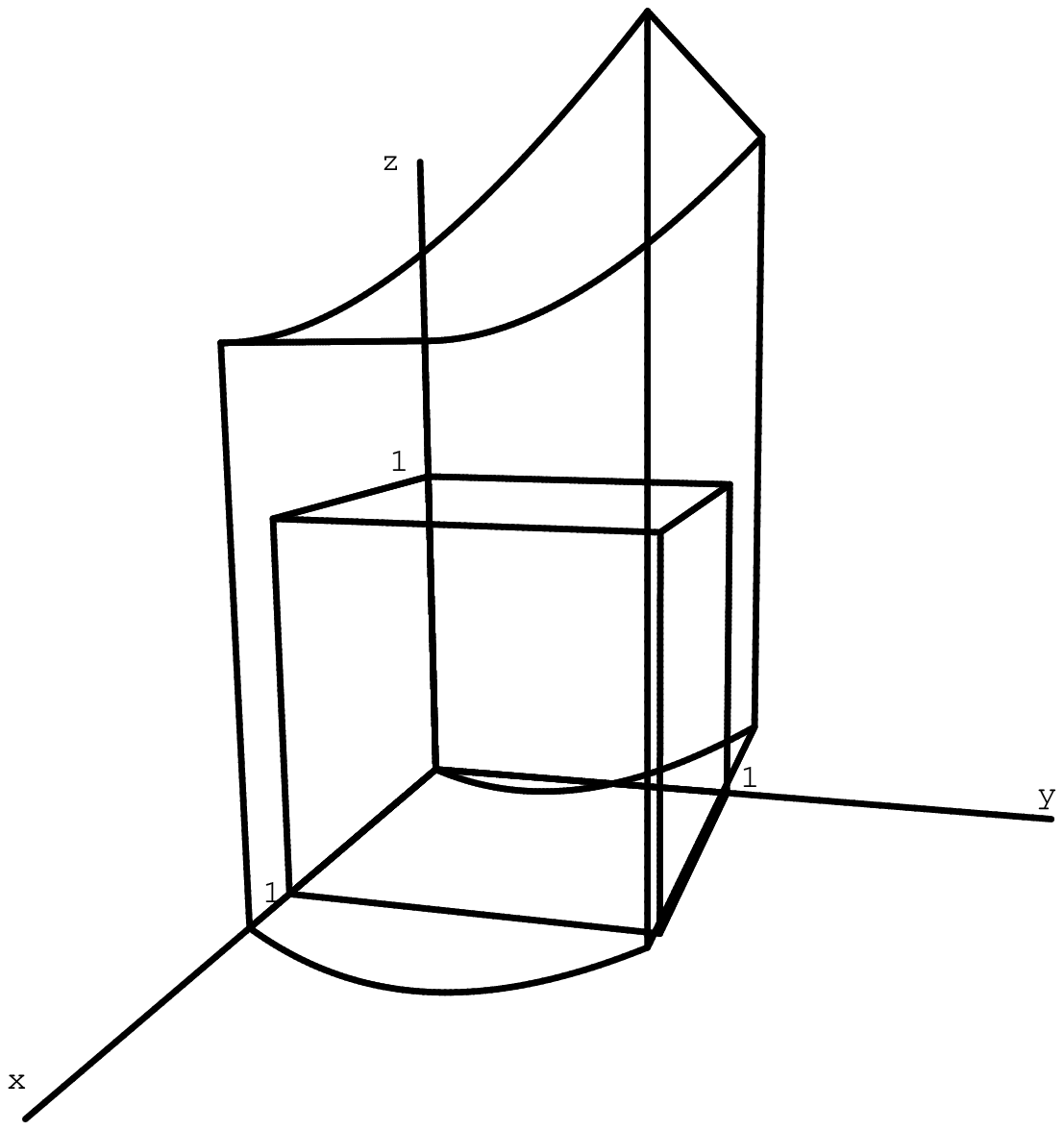}
\vfill
FIGURE 1
  \end{center}
\end{figure}

\eject

\begin{figure}[t]
\begin{center}
    \leavevmode
\epsfysize=7cm
\epsfbox{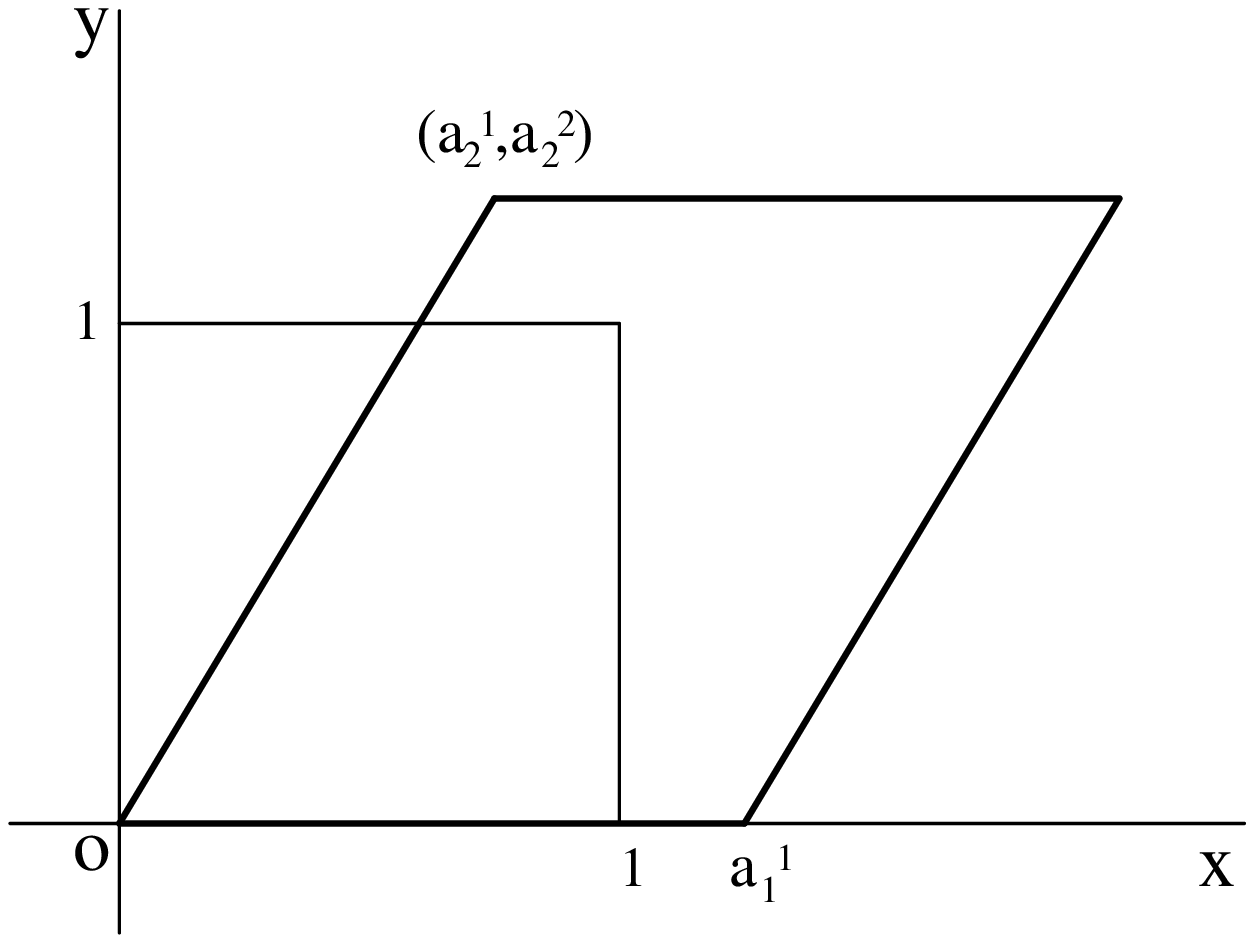}
\vfill
FIGURE 2
\end{center}
\end{figure}

\end{document}